\documentstyle[twoside,fleqn,espcrc2]{article}
\input epsf
\def \b{{\cal B}}
\def \beq{\begin{equation}}
\def \beqn{\begin{eqnarray}}

\def \cn{Collaboration}
\def \eeq{\end{equation}}
\def \eeqn{\end{eqnarray}}

\def \ite{{\it et al.}}

\def \st{\sqrt{3}}
\def \sx{\sqrt{6}}

\newcommand{\AmS}{{\protect\the\textfont2
  A\kern-.1667em\lower.5ex\hbox{M}\kern-.125emS}}

\title{Lattice QCD and heavy quarks}

\author{Jonathan L. Rosner \address{Enrico Fermi Institute and Department
        of Physics, University of Chicago \\ 
        5640 S. Ellis Avenue, Chicago, IL 60637 USA}}
               
\begin{document}

\begin{abstract}
Some interesting heavy-flavor questions are mentioned that may yield to the
methods of lattice QCD.  Special emphasis is placed on topics which arise in
the discussion of CP violation in $B$ decays.  Other subjects include
quarkonium and light-quark systems and some potential applications outside QCD.
\end{abstract}

\maketitle

\section{Introduction}

Quarks are bound into hadrons by the interaction of quantum chromodynamics
(QCD), some aspects of which cannot be treated by perturbation theory.  As a
result, non-perturbative methods have been developed, of which lattice gauge
theory is at present the leading contender. 

The study of hadronic properties of heavy-quark systems is valuable for at
least two reasons.  (1) By peeling away effects of the strong interactions, one
can uncover fundamental quark properties, such as the sources of masses, flavor
mixings, and CP violation as encoded in the Cabibbo-Kobayashi-Maskawa (CKM)
\cite{CKM} matrix.  (2) By drawing an analogy between heavy quarks and atomic
nuclei, and between light quarks and gluons and atomic electrons and
electromagnetic field quanta, one can simplify the description of hadrons. 
Systems with one heavy quark are like hydrogen atoms; the replacement of one
heavy quark by another is analogous to an isotope change.

I wish to touch on some aspects of (mostly) heavy-quark physics for which
lattice gauge theory can provide insights.  The conference organizers
originally called this talk ``Interesting Heavy Flavor Physics That Lattice
People Should Study,'' a provocative and peremptory title which seems to have
evoked the desired response \cite{GM}.

I will begin with a brief review of flavor-changing transitions among quarks as
described by the CKM matrix (Sec.~2).  Some questions demanding answers from
non-perturbative methods in QCD arise in the study of CP-violating decays of
$B$ mesons (Sec.~3).  Specific places where the lattice (or other approaches)
can help are noted in Sec.~4 for CP studies and in Sec.~5 for other
non-perturbative questions in heavy-quark physics.

Systems with more than one heavy quark are also worthy test-beds for methods in
QCD.  Bound states of heavy quarks and antiquarks (``quarkonium'') have yielded
information even when studied with the simplest nonrelativistic methods, but
questions remain, some of which the lattice seems uniquely poised to answer
(Sec.~6).  I would also like to mention a few favorite light-quark questions
(Sec.~7), and some areas outside QCD where the lattice could be of help
(Sec.~8).  A summary is contained in Sec.~9. 

\section{Quarks and the CKM Matrix}

We begin by updating some previous analyses \cite{JRKEK,JRUCLA,AJB,AK} in the
light of results from the summer 1998 conferences. The weak charge-changing
transitions between the quarks $i = (u,c,t)$ of charge 2/3 and those $j =
(d,s,b)$ of charge $-1/3$ are described by a unitary $3 \times 3$ matrix
$V_{ij}$ $(V^{-1} = V^\dag)$ which can be parametrized \cite{WP} as 
$$
V = \left[ \begin{array}{c c c}
1 - \frac{\lambda^2}{2} & \lambda & A \lambda^3 (\rho - i \eta) \\
- \lambda & 1 - \frac{\lambda^2}{2} & A \lambda^2 \\
A \lambda^3 (1 - \rho - i \eta) & -A \lambda^2 & 1 \\
\end{array} \right]~.
$$
Here $\lambda = \sin \theta_C \simeq 0.22$ is determined from strange particle
decays (for a recent analysis, see \cite{FJM}); $A \lambda^2 = V_{cb} = 0.0392
\pm 0.0027$ is obtained from decays of hadrons containing $b$ quarks to charmed
particles \cite{KB}, and $|V_{ub}| = (3.56 \pm 0.22 \pm 0.28 \pm 0.43)
\times 10^{-3}$ is the result of an average \cite{Rosnet} of several results
for decays of hadrons containing $b$ quarks to charmless final states. This
last result implies $|V_{ub}/V_{cb}| = 0.091 \pm 0.016$, or $(\rho^2 +
\eta^2)^{1/2} = 0.41 \pm 0.07$.  However, we still do not know $\gamma = {\rm
Arg}(V^*_{ub})$ well. 

Our uncertainty can be expressed as a region of allowed parameters in the
complex $\rho + i \eta$ plane.  The relation $V^*_{ub} + V_{td} = A \lambda^3$
is a consequence of the unitarity of the CKM matrix, so that a figure in the
complex plane with vertices $(\rho,\eta)$ (interior angle $\alpha$), $(1,0)$
(interior angle $\beta$), and $(0,0)$ (interior angle $\gamma$) is often
referred to as the {\it unitarity triangle} (see Fig.~1). The constraint on
$|V_{ub}/V_{cb}|$ then leads to an allowed annulus centered on (0,0) in the
$(\rho,\eta)$ plane. 

\begin{figure}
\centerline{\epsfysize = 1.3in \epsffile {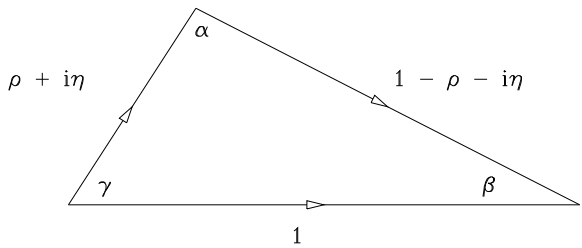}}
\caption{Unitarity triangle for CKM elements.  Here $\rho + i \eta =
V^*_{ub}/A \lambda^3$; $1 - \rho - i \eta = V_{td}/A \lambda^3$.} 
\end{figure}

CP-violating $K$-$\bar K$ mixing is encoded in the parameter $\epsilon_K
= (2.28 \pm 0.02) \times 10^{-3}$ \cite{PDG}.  In the CKM theory $\epsilon_K$
is due primarily to top-quark loops in the box diagrams governing mixing, and
so should be proportional to Im($V_{td}^2) \sim \eta(1-\rho)$.  Including the
contribution of charmed quarks in the loop, one can write the constraint
\cite{JRKEK,JRUCLA} as
\beq \label{eqn:Kmixcon}
\eta(1 - \rho + 0.44) = 0.51 \pm 0.18~~~.
\eeq
This relation defines a band bounded by two hyperbolae in the $(\rho,\eta)$
plane.  The error in (\ref{eqn:Kmixcon}) is dominated by that in $V_{cb}$;
we have used a parameter $\hat B_K = 0.8 \pm 0.2$ (the hat defines a particular
renormalization scheme) describing the matrix element of the short-distance
mixing operator between $K^0$ and $\bar K^0$. The error on the top quark's mass
\cite{PDG} is insignificant by comparison. 

The top quark dominates the loop diagrams governing $B^0$--$\bar B^0$ mixing. 
We have used a matrix element parameter $f_B \sqrt{B_B} = 200 \pm 40$ MeV to
extract a value of $|V_{td}|$ implying 
\beq \label{eqn:Bmixcon}
|1 - \rho - i \eta | = 1.01 \pm 0.21~~~.
\eeq  
This relation defines a $(\rho,\eta)$ region bounded by two circles with
centers at $(1,0)$.

A final constraint is provided by a new bound on mixing between the strange $B$
meson $B_s \equiv \bar b s$ and its antiparticle.  The mixing amplitude can be
parameterized in terms of the splitting between mass eigenstates:  $\Delta m_s
> 12.4$ ps$^{-1}$ (95\% c.l.) \cite{LEPBOSC}.  By comparing this value with the
corresponding one for non-strange $B$'s, $\Delta m_d = 0.471 \pm 0.016$
\cite{JA}, and using the estimate \cite{FM} $f_{B_s} \sqrt{B_{B_s}}/
f_B \sqrt{B_B} \le 1.25$ (see other talks in this Conference), one concludes
$|V_{ts}/V_{td}| > 4.0$ or $|1 - \rho - i \eta| < 1.14$.

\begin{figure}
\centerline{\epsfysize = 1.8in \epsffile {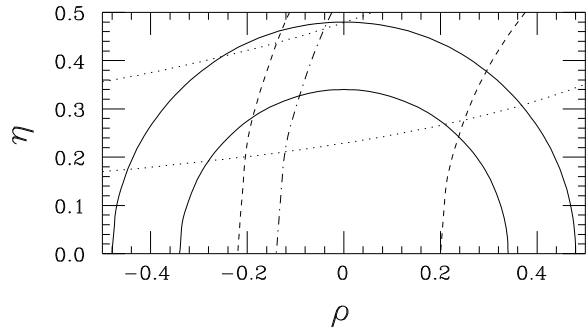}}
\caption{Region in the $(\rho,\eta)$ plane allowed by constraints on
$|V_{ub}/V_{cb}|$ (solid semicircles), $B^0$--$\bar B^0$ mixing (dashed
semicircles), CP-violating $K$--$\bar K$ mixing (dotted hyperbolae),
and $B_s^0$--$\bar B_s^0$ mixing (to the right of the dot-dashed semicircle).}
\end{figure}

The resulting constraints are shown in Fig.~2.  [The region of parameters is
slightly smaller than actually shown at the Conference as a result of
improvements in bounds on several parameters.]  The improved lower bound on
$\Delta m_s$ has contributed to the the evidence for $\eta \ne 0$ (i.e., for a
non-trivial phase in the CKM matrix) {\it independent of that provided by
CP-violating $K^0$--$\bar K^0$ mixing}.  The maximum allowed value of $\Delta
m_s$ allowed by this plot is about $(1.14/0.8)^2 \times 12.4~{\rm ps}^{-1}
\simeq 25~{\rm ps}^{-1}$, or $x_s \equiv \Delta m_s/\Gamma_s \simeq 40$.

The corresponding range of the angles $\alpha$, $\beta$, and $\gamma$, which
can be probed in $B$ decays (Sec.~3), are shown in Table 1. These correspond to
$-0.72 \le \sin 2 \alpha \le 0.90$, $0.55 \le \sin 2 \beta \le 0.85$, $0.54 \le
\sin^2 \gamma \le 1$. (For a slightly different analysis see \cite{Parodi}.
These authors, in our opinion, underestimate the errors on several key
quantitites such as $|V_{cb}|$ and obtain an allowed region which is a subset
of ours.)

\begin{table}
\caption{Range of angles of the unitarity triangle (Fig.~1) permitted by the
constraints in Fig.~2.}
\begin{center}
\begin{tabular}{r r r r r} \hline
Angle    &     & Deg & $\rho$  & $\eta$ \\ \hline
$\alpha$ & Min &  58 & $-0.09$ & 0.33   \\
         & Max & 113 & $ 0.25$ & 0.27   \\
$\beta$  & Min &  17 & $-0.09$ & 0.33   \\
         & Max &  29 & $ 0.22$ & 0.43   \\
$\gamma$ & Min &  47 & $ 0.25$ & 0.27   \\
         & Max & 105 & $-0.09$ & 0.33   \\ \hline
\end{tabular}
\end{center}
\end{table}

Many useful parameters contributing to this plot have been calculated or are
being refined in lattice QCD.  These include $B_K$, $f_B$, and $B_B$.  It is
quite likely, for example, that with an actual measurement (rather than a
bound) for $\Delta m_s$, and a good calculation of $f_{B_s} \sqrt{B_{B_s}}/
f_B \sqrt{B_B}$, one will be able to strengthen the case, already suggestive,
for nonzero $\eta$.

As an exercise in ``futurism,'' one can imagine a $(\rho,\eta)$ plot as shown
in Fig.~3 emerging in five years \cite{NSF}.  The potential for inconsistency
among all these measurements (pointing to new physics) is of course much
increased.  However, the constraints in Fig.~3 (aside from that due to
$\epsilon_K$) will largely circumvent any dependence on lattice or other
nonperturbative QCD calculations.  Hence, although lattice QCD has made great
strides in recent years, its days for ``prediction'' of certain quantities such
as $f_B$ may be numbered.  There will still be a need for others, such as
the ratio $f_{B_s} \sqrt{B_{B_s}}/f_B \sqrt{B_B}$ (to interpret $\Delta m_s/
\Delta m_d$) and the quantity $B_B$ itself (to interpret $\Gamma(B^+ \to \tau^+
\nu_\tau)/\Delta m_d$).

\section{CP-Violating $B$ Decays}

\subsection{New modes and their implications}

The branching ratios for decays of $B$ mesons to several exclusive charmless
final states have been pinned down in the past couple of years, improving the
prospects for learning the phases of CKM matrix elements and for seeing CP
violation.

\begin{figure}
\centerline{\epsfysize = 1.8in \epsffile {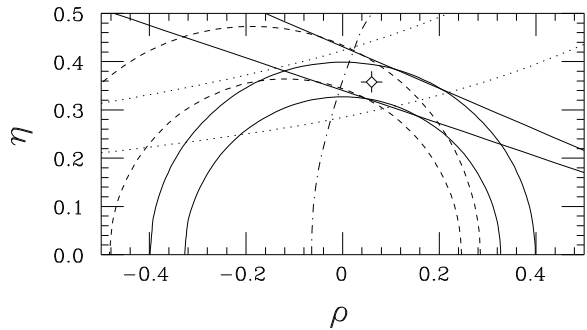}}
\caption{\small Example of a region in the $(\rho,\eta)$ plane that might be
allowed by data in the year 2003.  Constraints are based on the following
assumptions: $|V_{ub}/V_{cb}| = 0.08 \pm 0.008$ (solid semicircles),
$|V_{ub}/V_{td}|=|(\rho-i\eta)/(1-\rho-i\eta)| = 0.362 \pm 0.036$ based on
present data on $B^0$--$\bar B^0$ mixing and a measurement of $B(B^+ \to \tau^+
\nu_\tau)$ to $\pm 20\%$ (dashed semicircles), CP-violating $K$--$\bar K$
mixing as in Figure 2 except with $V_{cb}$ measured to $\pm 4\%$ (dotted
hyperbolae), the bound $x_s > 20$ for $B_s^0$--$\bar B_s^0$ mixing (to the
right of the dot-dashed semicircle), and measurement of $\sin 2 \beta$ to $\pm
0.059$ (diagonal straight lines).  The plotted point, corresponding to
$(\rho,\eta) = (0.06,0.36)$, lies roughly within the center of the allowed
region.} 
\end{figure}

The decay $B^+ \to K^0 \pi^+$ is expected to be a pure penguin process, while
$B^0 \to K^+ \pi^-$ and $B^+ \to K^+ \pi^0$, though dominated by penguin
amplitudes, should have other contributions at the level of some tens of
percent.  By comparing rates \cite{JA} for these processes, all of which have
branching ratios between one and two parts in $10^5$, one can learn about the
relative weak and strong phases of these various contributions; in particular,
one learns about the angle $\gamma$ illustrated in Fig.~1 \cite{RF,GR,NR}. 

The branching ratios for the decays $B^{+,0} \to K^{+,0} \eta'$ are quite large
\cite{CLEOetap} (averaging to $(6.8 \pm 1.1) \times 10^{-5}$ if they are equal
as one expects from penguin dominance of the decays).  The $\eta'$,
predominantly a singlet of flavor SU(3), appears to couple very favorably to
the rest of the system, whether due to an intrinsic gluonic or $c \bar c$
component or to the QCD anomaly.  One consequence of this enhanced coupling is
an improved prospect for seeing CP violation through unequal rates for $B^+ \to
\pi^+ \eta'$ and $B^- \to \pi^- \eta'$, while the amplitudes in $B^\pm \to
\pi^\pm \eta$ also turn out to favor CP-violating rate differences \cite{DGR}. 

Finally, $B$ decays to charmless final states involving one or two vector
mesons, such as $B^+ \to \pi^+ \omega$ and $B^+ \to K^+ \omega$ \cite{CLEOVP},
provide details of form factors which check specific dynamical models on which
many expectations are based \cite{VP}. 

\subsection{Rescattering issues}

The assumption that the decay $B^+ \to K^0 \pi^+$ is purely a penguin process
is called into question if rescattering from other final states, such as $K^+
\pi^0$, is important \cite{RF,Rescatt}.  If more than one amplitude contributes
to $B^+ \to K^0 \pi^+$, there can be an observable rate difference between that
process and its charge conjugate, and determinations of $\gamma$ by comparison
with other rates \cite{RF,GR,NR} are no longer so straightforward. 

In the context of a flavor-SU(3) analysis \cite{GHLR}, one normally expects
suppression of an ``annihilation'' amplitude involving the spectator quark in
which $\bar b u \to W^{*+} \to \bar s u \to (\bar s d)(\bar d u)$. 
Rescattering (e.g., through the $K^+ \pi^0$ state) can imitate such effects.
Present estimates of such contributions are very model-dependent.  However, a
flavor-SU(3) relation between rescattering contributions to $B^+ \to K^0 \pi^+$
and contributions to $B^+ \to \bar K^0 K^+$ can be obtained by the ``$U$-spin''
interchange $s \leftrightarrow d$ \cite{RF,Uspin}, with the result that the
rescattering contributions to $B^+ \to \bar K^0 K^+$ should be about
$|V_{ud}/V_{us}|^2 \simeq 20$ times those in $B^+ \to K^0 \pi^+$, and thus
should be large enough to enhance the decay rate of $B^+ \to \bar K^0 K^+$ by a
visible amount if they are at all important in $B^+ \to K^0 \pi^+$.

If a few intermediate states dominate the rescattering, one also expects a
visible enhancement of the $B^0 \to K^+ K^-$ decay rate, which would normally
be expected to be due to the suppressed $\bar b d \to \bar u u \to (\bar u
s)(\bar s u)$ ``exchange'' subprocess \cite{RSPP}.  The experimental upper
bound ${\cal B}(B^0 \to K^+ K^-) < 2.4 \times 10^{-6}$ is considerably better
than that ${\cal B}(B^+ \to K^+ \bar K^0) < 9.3 \times 10^{-6}$ discussed in
Refs.~\cite{RF,Uspin}, and is expected to be a factor of 60 below present
limits if the hierarchy described in Ref.~\cite{GHLR} is correct. 

\subsection{CP asymmetry in $B \to J/\psi K_S$}

The CKM theory predicts that mixing and decay amplitudes interfere to give a
rate difference between $B^0 \to J/\psi K_S$ and $\bar B^0 \to J/\psi K_S$: 
\beqn 
\frac{\Gamma(B^0 \to J/\psi K_S) - \Gamma(\bar B^0 \to
J/\psi K_S)}{\Gamma(B^0 \to J/\psi K_S) + \Gamma(\bar B^0 \to J/\psi K_S)}
\nonumber \\
= - \frac{x_d}{1 + x_d^2} \sin2 \beta~~~, \nonumber
\eeqn
where $x_d \equiv \tau(B^0) \Delta m_d$. The flavor of the neutral $B$ is
that {\it at time of production}.  This must be obtained either by studying the
flavor of the ``other'' $B$ produced in association, or by a ``same-side
tagging'' method \cite{GNR} in which the sign of a pion produced not far from
the neutral $B$ in phase space signifies its flavor.  Using this ``same-side''
method, both the OPAL and CDF Collaborations have reported asymmetries whose
central values are larger than physically expected (probably as a result of
uncertainty in estimating tagging efficiencies) but which exclude some
region of negative $\sin 2 \beta$:
\beqn
\sin 2 \beta & = & 3.2^{~+1.8}_{~-2.0} ~ \pm 
0.5~~({\rm OPAL}~~\protect\cite{OPALas})~, \nonumber \\
\sin 2 \beta & = & 1.8 \pm 1.1 \pm 0.3~~({\rm
CDF}~~\protect\cite{CDFas})~~~. 
\eeqn
For example, the CDF result excludes values of $\sin 2 \beta$ less than
$-0.20$ at 95\% confidence level.

The same-side tagging method uses the fact that if a $b$ quark fragments into a
$\bar B^0 = b \bar d$ meson, a $d$ quark must be nearby in rapidity. If this
$d$ quark materializes into a charged pion, that pion must be a $\pi^-$.  The
correlation between $\bar B^0$ and $\pi^-$, and between $B^0$ and $\pi^+$,
follows both from fragmentation models and from resonances. Although the $B^*$
is too light to decay to $B \pi$, there is a family of ``$B^{**}$'' resonances
expected to lie several hundred MeV above the $B$, so the decays $B^{**-} \to
\bar B^{(*)0} \pi^-$ and $B^{**+} \to B^{(*)0} \pi^+$ are permitted. 

The CDF Collaboration can find a suitable pion for tagging the neutral $B$ in
2/3 of the observed $J/\psi K_S$ decays (a sample of about 200 in the 110
pb$^{-1}$ accumulated during Run I).  The tagging method is calibrated by
measuring $B^0$--$\bar B^0$ oscillations in $B \to D^{(*)} l \bar \nu_l$
decays \cite{CDFcal}.  The ``dilution factor'' is measured to be ${\cal D}_0
\equiv 2 P_0 - 1 = 0.181^{+0.036}_{-0.032}$, where $P_0$ is the probability
that the tag correctly identifies the $B^0$ flavor. 

\section{Where the Lattice Can Help}

\subsection{Extraction of CKM matrix elements}

One can extract the CKM matrix elements $V_{cb}$ and $V_{ub}$ either from
exclusive \cite{WiseUCLA} or inclusive \cite{FalkUCLA} measurements. In either
case one needs theoretical guidance in passing from data at the hadron level
to conclusions at the quark level.

Lattice or other nonperturbative schemes can predict form factors for decays
such as $B \to D^{(*)} l \nu, ~\pi l \nu,~\rho l \nu, \ldots$ and $D
\to K^{(*)} l \nu,~\pi l \nu,~\rho l \nu, \ldots$.  Now, heavy-quark
methods permit one to use information in processes involving $D$ decays to pin
down hadronic effects in certain kinematic regions of $B$ decays.  Thus, one
can in principle avoid having to rely on theoretical form factor estimates.
However, until measurements of such rare processes as $B \to \rho l \nu$ and
$D \to \rho l \nu$ have reached the requisite accuracy, such estimates are
very useful.

Inclusive determinations of CKM matrix elements rely on comparison of data on
$B \to X_c l \nu$ and $B \to X_u l \nu$ with model predictions.  One has
to distinguish charmed inclusive states $X_c$ from non-charmed ones $X_u$ by
means of various kinematic variables, such as $M(X)$, lepton spectra, and
missing energy carried away by the neutrino.  Calculations of these variables
at the quark level give a zeroth-order approximation; for example, leptons
beyond the endpoint for $B \to X_c l \nu$ are assumed to have come from $B
\to X_u l \nu$.  A systematic expansion of differential and integrated decay
rates in inverse powers of the heavy quark mass exhibits our ignorance in terms
of a few parameters, which can either be extracted from data or calculated
using nonperturbative (e.g., lattice) methods. 

\subsection{Decay constants}

We have already noted the importance of the decay constant $f_B$ in determining
$|V_{td}|$ from $B^0$--$\bar B^0$ mixing, or $|V_{ub}|$ from $B^+ \to \tau^+
\nu$.  Lattice calculations appear to be the front-runners in estimating these
quantities.  As one example, we quote published results of the MILC
Collaboration \cite{MILC}:
\beqn
f_D     & = & 195 \pm 11^{+15+15}_{-8-0}~{\rm MeV}~~, \nonumber \\
f_{D_s} & = & 213 \pm 9^{+23+17}_{-9-0}~{\rm MeV}~~, \nonumber \\
f_B     & = & 159 \pm 11^{+22+21}_{-9-0}~{\rm MeV}~~, \nonumber \\
f_{B_s} & = & 175 \pm 10^{+28+25}_{-10-1}~{\rm MeV}~~~,
\eeqn
where the errors are statistical, systematic, and an estimate of the effects of
quenching, respectively.  The JLQCD Collaboration \cite{JLQCD} has found $f_D =
197 \pm 2$ MeV, $f_{D_s} = 224 \pm 2$ MeV, $f_B = 173 \pm 4$ MeV, and $f_{B_s}
= 199 \pm 3$ MeV in a quenched calculation.  The errors are statistical;
additional systematic and scale errors of 5\% (each) are estimated. The
observation of the decays $D_s \to \mu \nu$ and $D_s \to \tau \nu$ has
permitted the measurement $f_{D_s} = 255 \pm 21 \pm 28$ MeV \cite{Stone}, in
accord with these predictions. 

For a full review of heavy meson decay constants, see \cite{TD}.  Some salient
averages relevant to the physics of CKM matrix elements are \cite{GM,TD} $f_B
\sqrt{B_B} = 200 \pm 50$ MeV (a slightly more conservative error than we used
in Sec.~2), and $f_{B_s} \sqrt{B_{B_s}}/(f_B \sqrt{B_B}) = 1.17 \pm 0.06 \pm
0.12$ (again, slightly conservative compared to our quark-model estimate of
less than 1.25 for this ratio). 

\subsection{Spectroscopy of orbitally excited mesons}

The spectroscopy of P-wave levels of a heavy quark (e.g., $\bar b$) and a light
quark (e.g., $u$) can be of interest for tagging the flavor of neutral $B$
mesons.  Moreover, these $B^{**}$ levels (and their lighter relatives $D^{**}$
involving charmed quarks) are intrinsically interesting as tests of theories of
the spectrum.

One describes P-wave bound states of a single heavy antiquark $\bar Q$ and a
light quark $q$ in the following manner \cite{HQ}.  First, couple the light
quark's spin ($s=1/2$) to its orbital angular momentum ($l=1$) to form total
light-quark angular momenta $j = 1/2$ or $j = 3/2$. Then, couple $j$ to the
heavy-quark spin to form total angular momenta $J = 0$, 1 (twice), and 2. 
Labelling states by the notation $J^P_j$, where the parity $P$ is even for
$l=1$ $\bar Q q$ states, we then have states $0^+_{1/2},~1^+_{1/2},~1^+_{3/2}$,
and $2^+_{3/2}$.  The splittings between $0^+_{1/2}$ and $1^+_{1/2}$ levels,
and between $1^+_{3/2}$ and $2^+_{3/2}$ levels, should be of order $1/m_Q$.

The $j=1/2$ states are predicted to decay to the ground $\bar Q q$ states and a
pion in an S-wave, while the $j=3/2$ states should undergo these decays in a
D-wave and hence should be considerably narrower.  It is likely that these
latter ones are the states that have been seen in the $D^{**}$ and $B^{**}$
systems. 

Predictions of the masses and decay widths of the unseen $j=1/2$ states are
thus of great interest.  Do they contribute in a significant way to the $\pi
B^{(*)}$ correlations useful in same-side tagging?  At this Conference, some
encouraging lattice results have been presented \cite{AAK} indicating that the
$0^+_{1/2}$ level lies about 130 MeV below the $2^+_{3/2}$ level, and that the
$0^+_{1/2}$--$1^+_{1/2}$ splitting is of order $1/m_Q$ as expected. 

Another point of interest is the nature of $D^{**}$ resonances that contribute
to the roughly 1/3 of all semileptonic $B \to X_c l \nu$ decays not involving
$X_c = D$ or $D^*$ \cite{CLEOsl}.  Such resonances decay to $D \pi$ (for $J^P =
0^+,~2^+$) or $D^* \pi$ (for $J^P = 1^+,~2^+$). If the semileptonic decay $B
\to D^{(*)} l \nu$ also involves another ``secondary'' pion, that pion can be
confused with the ``same-side'' tagging pion if vertex resolution is poor.  The
sign of the secondary pion is opposite to that of the same-side tagging pion
for a given flavor of $B$, so good understanding of secondary pion production
in $B$ semileptonic decay is highly desirable \cite{CDFcal,Maks}. 

\subsection{Rescattering questions}

One may be asking too much for the lattice to estimate final-state
interaction effects in low-multiplicity decays of $D$ or $B$ mesons.  These
effects, in $B$ decays, are crucial in interpreting certain varieties of
CP-violating asymmetries, should they arise in future data, in terms of
fundamental CKM phases.  As two examples, we pose the following. 

(1) Naive estimates indicate that the decay rate for $B^0 \to K^+ K^-$ should
be very small in comparison with some related modes (such as $B^+ \to K^+
\bar K^0$).  This is because $B^0 \to K^+ K^-$ either requires the $\bar b$ and
$d$ in the $B^0$ to exchange a $W$ and materialize into $\bar u u$, or to
proceed via rescattering from some less-suppressed intermediate state.  Can
the lattice say anything about this process?

(2) Final-state phase differences in $B \to K \pi$ scattering have been
estimated in some quarters to be small \cite{Rescatt}.  This is a pity as
otherwise conditions could be favorable for a large CP-violating difference
between the rates for such processes as $B^0 \to K^+ \pi^-$ or $B^+ \to
K^+ \pi^0$ and their charge conjugates.  If there is a source of large
final-state phases, it could be the charm-anticharm intermediate state
\cite{ChPen}.  Can the lattice say anything about this?

\section{Other Nonpertubative Questions}

\subsection{The $D^* D \pi$ coupling constant}

It is possible to calculate strong coupling constants in lattice gauge theories
\cite{sc}.  The $D^* D \pi$ coupling is of interest for a couple of reasons.

1) The hadronic widths of $D^*$ states are too small to be measured directly.
All that exists is an upper limit \cite{ACCMOR} $\Gamma_{\rm tot}(D^{*+})
< 130$ keV. However, the $D^{*0}$ branching ratios for hadronic and
electromagnetic decays are comparable to one another.  Now, the $D^{*0} \to D^0
\gamma$ width depends on the magnetic transition moment of the charmed quark
(which is calculable) and that of the $\bar u$ quark (which is much harder to
estimate). These two contributions interfere constructively in the matrix
element. 

Recently the CLEO Collaboration \cite{CLEOdg} has measured the ratio $\b(D^{*+}
\to D^+ \gamma) = (1.68 \pm 0.42 \pm 0.29 \pm 0.03)\%$, where the first error
is statistical, the second is systematic, and the third is associated with
uncertainty in a kinematic factor describing the ratio of the $D^0 \pi^+$ and
$D^+ \pi^0$ decays.  This is to be contrasted with the much larger value
\cite{PDG} of $\b(D^{*0} \to D^0 \gamma) = (38.1 \pm 2.9)\%$ The branching
ratio of $D^{*+} \to D^+ \gamma$ is so small because the contributions of the
charmed and light quark interfere destructively, almost cancelling one another.
By combining this information with hadronic branching ratios, including
$\b(D^{*+} \to D^+ \pi^0) = (30.73 \pm 0.13 \pm 0.09 \pm 0.61)\%$ and
$\b(D^{*+} \to D^0 \pi^+) = (67.59 \pm 0.29 \pm 0.20 \pm 0.61)\%$ \cite{CLEOdg}
and $\b(D^{*0} \to D^0 \pi^0) = (61.9 \pm 2.9)\%$ \cite{PDG}, one can solve
for the light-quark transition moment \cite{AmD}, thereby calibrating all the
$D^{*+}$ and $D^{*0}$ widths absolutely.  One finds $\Gamma_{\rm tot}(D^{*+}) =
90^{+50}_{-30}$ keV, in accord with the ACCMOR limit. Defining the $D^*D \pi$
coupling $g$ to be 1 in the constituent-quark limit, this range of widths
implies $g = 0.56 \pm 0.11$ in the treatment of \cite{AmD}, or $g =
0.27^{+0.04+0.05}_{-0.02-0.02}$ taking account of higher-order terms in chiral
perturbation theory \cite{Stewart}.

2) The $D^* D \pi$ coupling is relevant to some estimates of hadronic effects
in semileptonic $B$ decays.  The coupling constant $g$ enters into part of the
nonperturbative ${\cal O}(1/m_c^2)$ corrections to the Isgur-Wise $B \to D^*$
form factor ${\cal F}$ at the zero-recoil point, which affects the
determination of $V_{cb}$ \cite{WiseUCLA}. 

\subsection{Lifetime hierarchies}

Although the lifetimes of charmed hadrons vary by factors of more than 10,
from less than 0.1 ps for the $\Omega_c$ to greater than 1 ps for the $D^+$
\cite{PDG}, conventional wisdom \cite{NS} predicts less than a 10\% variation
among hadrons containing $b$ quarks.  The fact that the lifetime of the
$\Lambda_b$ is only about 0.8 times that of the $B^{+,0}$ and $B_s$ mesons is
not understood at present (see, e.g., \cite{NS,LB}).  Some non-perturbative
effects, not accounted for by the usual estimates, are apparently at work.
The lattice may be able to shed some light on this question.

\subsection{Branching ratios of $\Lambda_c$}

There exist no direct measurements of $\Lambda_c$ branching ratios.  The
absence of this information has far-reaching consequences on estimates
of $\Lambda_c$ production and other ``engineering'' quantities \cite{DLC}.
Branching ratios are calibrated by assuming that the exclusive semileptonic
decay $\Lambda_c \to \Lambda l \nu$ saturates the inclusive semileptonic rate,
which is then calculated perturbatively.  It would be helpful if the lattice
or some other nonperturbative scheme could provide form factors for $\Lambda_c
\to \Lambda$ transitions, and guidance about what other states (if any) are
likely to be excited.

\subsection{$\Sigma_b^{(*)}$ spectra}

A couple of years ago the DELPHI Collaboration \cite{DELS} claimed a large
splitting $M(\Sigma_b^*) - M(\Sigma_b) = 56 \pm 16$ MeV.  This value is hard to
understand on the basis of heavy-quark physics \cite{FB}.  In the states
$\Sigma_Q$ and $\Sigma^*_Q$, where $Q$ is a heavy quark, the light quarks $q$
are coupled to a state of isospin $I_{qq} = 1$ and spin $S_{qq} = 1$.  This
spin is then coupled to the spin $S_q$ of the heavy quark $Q$ to give total
angular momentum $J = 1/2$ for $\Sigma_Q$ or $J = 3/2$ for $\Sigma^*_Q$.  The
hyperfine splitting between these two states should be inversely proportional
to the heavy quark mass $m_Q$, so that one expects $M(\Sigma_b^*) - M(\Sigma_b)
= (m_c/m_b) [M(\Sigma_c^*) - M(\Sigma_c)] \simeq (1/3)[M(\Sigma_c^*) -
M(\Sigma_c)]$. 

Since the DELPHI report first appeared, the CLEO Collaboration has presented
convincing evidence for the $\Sigma^*_c$ at a mass about 65 MeV above the
$\Sigma_c$ \cite{CLEOSg}.  One would then expect $M(\Sigma_b^*) - M(\Sigma_b)$
to be around 20 MeV, lower than the DELPHI result and in accord with a lattice
estimate presented at this Conference \cite{AAK}. 

\section{Quarkonium Issues}

The bound states of a heavy quark $Q$ and the corresponding antiquark $\bar Q$,
known as quarkonium, have contributed much to our understanding of QCD.  The
lattice has used these systems to extract remarkably precise values of the
strong coupling constant and to study the behavior of light degrees of freedom
(quark-antiquark pairs and gluons) surrounding the nonrelativistic $Q \bar Q$
system \cite{QAS}.  Some other possible topics of interest are mentioned below.

\subsection{Universal separation corresponding to flavor threshold}

The number of $Q \bar Q$ bound states below flavor threshold can be shown to
increase as $m_Q^{-1/2}$ \cite{EG,QRn}.  This result is easily seen using a WKB
estimate \cite{QRn} and the assumption of a universal $Q \bar Q$ separation
corresponding to flavor threshold.  If $M(Q \bar q)$ denotes the mass of the
lightest flavored meson, the $Q \bar Q$ potential at threshold separation
$r_{\rm th}$ should satisfy $V(r_{\rm th}) = 2 M(Q \bar q) - 2 m_Q$, and as
$m_Q \to \infty$ this quantity should approach a constant.  An estimate of
the corresponding value of $r_{\rm th}$ \cite{JRth} is 1.4 to 1.5 fm.  This
appears to be in the range of lattice estimates \cite{Latth}, and studies are
continuing \cite{HW}.

\subsection{Mixing of S and D states}

Firm predictions exist for the masses of D-wave quarkonium states
\cite{Kwong} (if they are not perturbed by nearby thresholds).
Such states are difficult to observe.  They can be produced via electromagnetic
transitions from P-wave levels, or in $e^+ e^-$ collisions via mixing with
S-wave states.  Estimates of this mixing vary.  Intermediate states consisting
of pairs of flavored mesons probably play a key role, especially for the D-wave
charmonium states, all but two of which are almost certain to lie above flavor
threshold.  The exceptions, the lowest-lying $^1D_2$ and $^3D_2$ levels, may
lie low enough in mass to forbid their strong decays, which cannot occur to $D
\bar D$ and must at least involve $D \bar D^*$ or $\bar D D^*$. Searches for
these last two levels are part of the program of charmonium studies in the
Fermilab Accumulator Ring \cite{Fccbar}.  Searches for the D-wave $b \bar b$
levels are possible at CESR if the energies for running below the
$\Upsilon(4S)$ -- a key component of background studies for $B$ production --
are chosen appropriately. 

Lattice estimates of S--D mixing thus would be very helpful.  Such estimates
probably must await a more thorough understanding of the role of light quark
pairs, which undoubtedly are a key feature of this mixing. 

\subsection{Masses and transition matrix elements of $\eta_b$ states}

The $^1S_0$ levels of the $b \bar b$ system -- the $\eta_b,~\eta_b', ~
\eta_b'',~\ldots$ states -- have not yet been seen.  Their masses influence
determinations of $\alpha_s$ from quarkonium spectroscopy \cite{QAS} because
one would like to use spin-averaged levels spacings (for example, in the
comparison of 1P -- 1S and 2S -- 1S spacings) but this is not possible as long
as spin-singlet levels have not been seen.  Thus, one must either work with the
observed spin-triplet levels or make a theoretical correction for the
spin-splittings. 

Estimates of hyperfine splittings between $^3S_1$ and $^1S_0$ based on
perturbation theory \cite{NPT} indicate that the next-to-leading-order
corrections are very important.  Moreover, the square $|\Psi(0)|^2$ of the
S-wave wave function at zero interquark separation -- intrinsically a
nonperturbative quantity -- enters such calculations.  Leptonic widths are
sensitive to $|\Psi(0)|^2$, but with important relativistic corrections --
again indicating the importance of methods transcending perturbation theory. It
appears that lattice methods have some difficulty in estimating heavy-quark
hyperfine splittings, but it is worth thinking whether some lattice insight
might nonetheless complement the more usual methods. 

The best prospects for producing the $\eta_b$ -- the $1^1S_0$ $b \bar b$ level
-- are probably through the transition $\Upsilon(2S) \to \gamma \eta_b$.
Although the wave functions of 2S and 1S states are orthogonal to one another,
two effects combine to give a non-zero transition amplitude.  First, the matrix
element must be taken of the spherical Bessel function $j_0(kr/2)$, where $k$
is the photon energy and $r$ is an interaction radius \cite{Bessel}.  When $kr$
is non-negligible, this matrix element will not vanish.  Second, hyperfine
interactions can distort the $^3S_1$ and $^1S_0$ wave functions in different
ways.  An estimate of the first effect leads to the estimate ${\cal
B}(\Upsilon(2S) \to \gamma \eta_b) \simeq 10^{-4}$ \cite{MS}.  It is
probably worth updating this estimate in the light of all the progress on
quarkonium in the past 15 years.  The photon in this transition, of energy
around 600 MeV, should be detectable with enough $e^+ e^-$ collisions at the
c.m. energy of the $\Upsilon(2S)$. 

\subsection{Hybrid states}

The lattice has been able to predict the masses of hybrid states composed of
both quarks and gluons \cite{hybrids}.  Hybrid states $Q \bar Q g$ have been
suggested recently \cite{IDhc} as a possible solution to the ``missing charm''
problem in $b$ decays.  One is looking for a mechanism whereby the decay $b \to
c \bar c s$ is enhanced but does not lead to final states with visible charm.
The known states below charm threshold apparently do not suffice.  If hybrid
states $c \bar c g$ above $D \bar D$ threshold are produced with an enhanced
rate (deemed unlikely in one calculation, however \cite{CFP}) and then decay
primarily via $c \bar c$ annihilation to gluonic and/or light-quark systems,
one may be able to understand why fewer charmed particles than predicted are
seen in $b$ decays, and why the branching ratio of $b$ quarks to nonleptonic
final states is slightly higher than expected.  To check this hypothesis
requires estimates of production, masses, and decay rates of hybrid mesons. 

\subsection{Exotic systems}

Although all known hadrons so far consist of a quark-antiquark pair or
three quarks, other stable color singlets may exist.  These are conventionally
known as ``exotic states.''  The presence of a heavy quark has been predicted
to stabilize such states.  Thus, for example, it has been predicted that
there are stable ``pentaquark'' states of the form $qqqs\bar c$ and $qqqs
\bar b$ \cite{pq}, where $q$ stands for $u$ or $d$.  A recent quark-model
estimate \cite{DBL} predicts that the lowest charmed states decay strongly,
while those with a $\bar b$ are stable except for weak decays.  What does
the lattice say?

\section{Some Light-Quark Questions}

\subsection{Exotic light-quark states}

Exotic states of light quarks have been predicted over the years in various
forms.  There is no consensus on their properties, 35 years after the
introduction of the quark model.  Open questions include the following: 

(1) Is a $\Lambda \Lambda$ dibaryon (the ``$H$'') a bound state?
Quark-model calculations \cite{Jaffe} indicate a gain in binding energy when
quark spins are recoupled to gain the maximum possible hyperfine attraction.
The two $\lambda$'s still bind in the presence of SU(3) breaking \cite{JRH}. 
However, instanton effects \cite{Nussinov} may invalidate this conclusion.

(2) Are there $K \bar K$ ``molecules,'' such as $f_0$ and $a_0$, near
threshold?  How about $K \bar K \pi$ molecules, e.g., to account for
the state $\eta(1410)$ \cite{JRMol} ?

(3) States of two quarks and two antiquarks were discussed at this Conference
\cite{Pen}.  Are there such states with strong couplings to baryon-antibaryon
pairs \cite{JRex} or to pairs of vector mesons \cite{RJex}?  One possible
explanation of the fact that $\sigma(\gamma \gamma \to \rho^0 \rho^0) \gg
\sigma(\gamma \gamma \to \rho^+ \rho^-)$ near threshold is the cooperation of
direct-channel resonances with $I=0$ and $I=2$ \cite{res}. 

\subsection{$\Lambda \pi$ phase shifts}

Can lattice QCD say anything about low-energy $\Lambda \pi$ scattering?  The
difference $\delta$ between S-wave and P-wave phase shifts at c.m. energy equal
to $M(\Xi)$ governs the size of observable CP-violating effects in comparison of
$\Xi^- \to \Lambda \pi^-$ and $\bar \Xi^+ \to \bar \Lambda \pi^+$ decays, the
subject of an experimental search at Fermilab \cite{HyperCP}.  Recent
calculations based on chiral perturbation theory \cite{SPdelta} find
values of $\delta \equiv \delta_S - \delta_P$ of order a few degrees, implying
small CP asymmetries.

\subsection{Weak decay matrix elements}

The application of lattice gauge theory to certain weak processes (for
instance, those involving kaons) has a long history \cite{GM,Ktwopi}. Lattice
methods might also be tried in radiative hyperon decays:  $\Sigma \to p
\gamma$, $\Lambda \to n \gamma$, $\Xi \to \Lambda \gamma$, $\Xi \to \Sigma
\gamma$, and $\Omega \to \Xi \gamma$. One seeks predictions of rates and of
parity-violating asymmetries. Experimental upper limits for another process,
the $|\Delta S| = 2$ decay $\Xi^0 \to p \pi^-$, have recently been lowered to
$\b < 1.7 \times 10^{-5}$ \cite{Kreutz}.  This is far above the standard model
range \cite{SMX}, which is not well known at present.  Perhaps the lattice can
help here. 

\subsection{Glue content of $\eta'$}

Before the large branching ratio for the decays $B \to K \eta'$ was discovered
\cite{CLEOetap}, it was proposed \cite{GReta} that the large flavor-singlet
component of the $\eta'$ could lead to a significant amplitude for this process
as a result of the two-gluon intermediate state in penguin processes.  A
related question is the glueball content of the $\eta'$.  The decay $\eta' \to
\rho \gamma$ and the recently observed process \cite{phi} $\phi \to \eta'
\gamma$ with branching ratio $\b = (1.35^{+0.55}_{-0.45}) \times 10^{-4}$
proceed at rates consistent with a ``normal'' $q \bar q$ content of $\eta'$
\cite{JLRetap}.  (See, however, the arguments advanced in \cite{BFT}.) A good
approximation to the $\eta$ and $\eta'$ wavefunctions corresponds to mixing
between flavor octet and singlet states $\eta_8 \equiv (u \bar u + d \bar d - 2
s \bar s)/\sx$ and $\eta_1 \equiv (u \bar u + d \bar d + s \bar s) /\st$ with
an angle of $\theta = \sin^{-1}(1/3) \simeq 19.5^{\circ}$: $\eta \simeq (u \bar
u + d \bar d - s \bar s)/\st$, $\eta' \simeq (u \bar u + d \bar d + 2 s \bar
s)/\sx$ \cite{JLRetap,etamix}. 

\section{Non-QCD Axes to Grind}

\subsection{Dynamical electroweak symmetry breaking}

Most effort in lattice gauge theory has been devoted to QCD, for which there is
overwhelming evidence from perturbative approaches. However, lattice methods
may also guide searches for theories which are not yet well-established, such
as dynamical electroweak symmetry breaking (``technicolor'' \cite{TC})
theories.  Non-perturbative methods based on low-energy theorems, crossing
symmetry, and unitarity have been used \cite{RR} to argue that whereas some
features of these schemes may be expected for a wide class of theories, others
may be more sensitive to details of models.  As one example, the minimal
technicolor model \cite{TC} based on fermions $F$ with charges $Q = \pm 1/2$
leads to an anomaly-free gauge sector.  Higgs and Nambu-Goldstone bosons in
this model are $F \bar F$ pairs.  The lattice should be able to provide
insights about the $F \bar F$ spectrum, depending on the underlying
interaction. 

\subsection{Composite models}

Composite models of quarks and leptons require a large mass hierarchy in which
the properties of very light states appear to be determined at a much higher
mass scale (since no evidence for deviations from pointlike structure has
appeared up to scales of several TeV), and the need to describe chiral
fermions.  This last feature is a particular obstacle to the application of
lattice methods, at least for the moment. 

An example of a simple model \cite{FS} in search of a theory is to imagine that
the fermions $F$ mentioned above are the only fermionic constituents of matter,
with quarks and leptons made of $F \bar S$ pairs, where $S$ are scalars which
are either color singlets with charges $Q = \pm 1/2$ (leading to leptons) or
color triplets with $Q = \pm 1/6$ (leading to quarks).  A question in this
picture (which will depend on the underlying dynamics) is the relative masses
of $F \bar F$, $F \bar S$, $S \bar S$, and baryonic (multi-$F$)  states.  Do we
want light $S \bar S$ states?  If not, are there theories where they are
naturally heavy?  Do $F \bar F$ condensates form, leading to the desired
$\Delta I = 1/2$ masses?  What about multi-$F$ condensates, which could lead to
large Majorana masses for right-handed neutrinos? 

\subsection{A phase transition to supersymmetry?}

The model just described has the potential for supersymmetry.  It involves an
isodoublet of Dirac fermions $F_{\pm}$, each with four components, a quadruplet
of scalars $S$ with charges $Q = -1/2,1/6,1/6,1/6$ (the last three values
referring to the three colors), and a corresponding quadruplet of antiscalars
$\bar S$ with charges $1/2,-1/6,-1/6,-1/6$.  There is an equal number of
fermionic and bosonic degrees of freedom but no $N=1$ supersymmetry since the
charges are different.  However, suppose there were a transition at high
temperature to a phase where the charges of the scalars were
$-1/2,-1/2,1/2,1/2$ and those of the antiscalars were $1/2,1/2,-1/2,-1/2$. If
so, color and electromagnetism would no longer commute; quarks would have
integer charges as in the Han-Nambu \cite{HN} model.  (A related transition at
high {\it density} was suggested at this conference by M. Alford \cite{MA}.)
Supersymmetry might be a desirable feature of a composite model by explaining
the presence of light fermions in the spectrum. 

\section{Summary}

The lattice has been shown to be a useful tool for peering beneath the
complexities of hadron physics to learn about fundamental properties of quarks.
It has permitted us to extend the usefulness of QCD beyond perturbation theory,
and to explore the strong-coupling behavior of quantum field theories other
than QCD.

In systems containing heavy quarks the lattice has given us a number of
results, but time is running out for ``predictions'' rather than merely
``postdictions.'' For example: 

1.  There has been substantial progress on heavy meson decay constants such
as $f_B$.  As a result, one hopes to be able to extract better limits on the
CKM matrix element $|V_{td}|$ from the observed strength of $B^0$--$\bar B^0$
mixing.  At present $f_B \sqrt{B_B}$ and hence $|V_{td}|$ are known to about
20\%.  The measurement of other quantities such as the $B_s$--$\bar B_s$ mixing
amplitude and the $B^+ \to \tau^+ \nu_\tau$ branching ratio may lead to more
precise information on $|V_{td}|$ before lattice calculations achieve much
greater accuracy. 

2.  Masses and widths of resonances containing a single heavy quark are
starting to be predicted by the lattice.  One is particularly interested in
those orbitally excited mesons which have not yet been discovered, since they
may play a role in the identification of neutral $B$ meson flavor at the time
of production.  However, experimental study of these mesons (and also of the
corresponding baryons) is proceeding apace.  Other questions about $D$ and $B$
mesons include the final states populated by their semileptonic decays, the
corresponding form factors for such decays, and the thorny problem of
non-leptonic decays. 

3.  Quarkonium systems generally present the lattice theorist with a set of
given data from which interesting quantities (such as $\alpha_s$) can be
extracted.  It would be interesting also to see some predictions for as yet
unseen states, such as the $^1$S$_0$ $b \bar b$ mesons (whose masses influence
those extractions of $\alpha_s$) and the D-wave $b \bar b$ systems (whose
masses we think we know well, but perhaps not their procudction cross sections
in $e^+ e^-$ annihilations).  Do heavy quarkonium systems really dissociate
into pairs of flavored mesons at a universal separation of 1.4 to 1.5 fm?  Both
data and the lattice seem to indicate so. 

Altogether it looks as if a wide range of problems may be accessible by lattice
methods.  Although lattice theorists have been attacking many of them, I hope
this talk has indicated at least some others worth trying.

\section{Acknowledgements}

I would like to thank Amol Dighe, Isi Dunietz, and Michael Gronau for enjoyable
collaborations on many of the subjects mentioned here, and Tom DeGrand for the
invitation to speak at this conference and for helpful advice.  Mark Alford,
Adam Falk, Sheldon Stone, and Mark Wise also provided welcome comments.  Some
of the research described here was performed at the Aspen Center for Physics. 
This work was supported in part by the United States Department of Energy under
Grant No.~DE FG02 90ER40560. 
\bigskip

\def \ajp#1#2#3{Am.~J.~Phys. #1 (#3) #2}
\def \ap#1#2#3{Ann.~Phys.~(N.Y.) #1 (#3) #2}
\def \apny#1#2#3{Ann.~Phys.~(N.Y.) #1 (#3) #2}
\def \app#1#2#3{Acta Physica Polonica #1 (#3) #2}
\def \arnps#1#2#3{Ann.~Rev.~Nucl.~Part.~Sci. #1 (#3) #2}
\def \arns#1#2#3{Ann.~Rev.~Nucl.~Sci. #1 (#3) #2}
\def \art{and references therein}
\def \ba88{Particles and Fields 3 (Proceedings of the 1988 Banff Summer
Institute on Particles and Fields), edited by A. N. Kamal and F. C. Khanna
(World Scientific, Singapore, 1989)}
\def \baps#1#2#3{Bull.~Am.~Phys.~Soc. #1 (#3) #2}
\def \be87{Proceedings of the Workshop on High Sensitivity Beauty
Physics at Fermilab, Fermilab, Nov. 11--14, 1987, edited by A. J. Slaughter,
N. Lockyer, and M. Schmidt (Fermilab, Batavia, IL, 1988)} 
\def \btasi{Testing the Standard Model (Proceedings of the 1990
Theoretical Advanced Study Institute in Elementary Particle Physics),
edited by M. Cveti\v{c} and P. Langacker (World Scientific, Singapore, 1991)}
\def \bucla{Beauty '97 (Proceedings of the Fifth International Workshop on
$B$-Physics at Hadron Machines, Los Angeles, USA, October 13--17, 1997),
edited by P. Schlein (North-Holland, Amsterdam, 1998), Nucl.~Inst.~Meth.~A
408 (1998)}
\def \cmts#1#2#3{Comments on Nucl.~and Part.~Phys. #1 (#3) #2}
\def \cn{Collaboration}
\def \corn{Lepton and Photon Interactions:  XVI International Symposium,
Ithaca, NY 1993, edited by P. Drell and D. Rubin (AIP, New York, 1994)}
\def \cp89{{\it CP Violation,} edited by C. Jarlskog (World Scientific,
Singapore, 1989)} 
\def \dpfa{The Albuquerque Meeting:  DPF 94 (Division of Particles and
Fields Meeting, American Physical Society, Albuquerque, NM, August 2--6,
1994), ed. by S. Seidel (World Scientific, River Edge, NJ, 1995)}
\def \dpff{The Fermilab Meeting -- DPF 92 (Division of Particles and
Fields Meeting, American Physical Society, Fermilab, 10--14 November, 1992),
ed. by C. H. Albright \ite~(World Scientific, Singapore, 1993)} 
\def \dpfm{The Minneapolis Meeting:  DPF 96 (Division of Particles and
Fields Meeting, American Physical Society, Minneapolis, MN, 10--15 August,
1996), to be published}
\def \dpfv{The Vancouver Meeting - Particles and Fields '91
(Division of Particles and Fields Meeting, American Physical Society,
Vancouver, Canada, Aug.~18--22, 1991), ed. by D. Axen, D. Bryman, and M. Comyn
(World Scientific, Singapore, 1992)} 
\def \efi{Enrico Fermi Institute Report No.~}
\def \epj#1#2#3{Eur.~Phys.~J.~#1 (#3) #2}
\def \fermlg{Proc.~Int.~Symp.~on Lepton and Photon Interactions at High
Energies (Fermilab, August 23--29, 1979), T. B. W. Kirk and H. D. I.
Abarbanel, eds., Fermilab, Batavia, IL (1979)}
\def \hb87{Proceeding of the 1987 International Symposium on Lepton and
Photon Interactions at High Energies, Hamburg, 1987, ed. by W. Bartel
and R. R\"uckl (Nucl.~Phys.~B, Proc. Suppl., vol. 3) (North-Holland,
Amsterdam, 1988)}
\def \ib{{\it ibid.}}
\def \ibj#1#2#3{{\it ibid.} #1 (#3) #2}
\def \ijmpa#1#2#3{Int.~J. Mod.~Phys.~A #1 (#3) #2}
\def \jpb#1#2#3{J. Phys.~B #1 (#3) #2}
\def \jpg#1#2#3{J. Phys.~G #1 (#3) #2}
\def \KEK{Flavor Physics (Proceedings of the Fourth International
Conference on Flavor Physics, KEK, Tsukuba, Japan, 29--31 October 1996),
edited by Y. Kuno and M. M. Nojiri, Nucl.~Phys.~B Proc.~Suppl.~59 (1997)}
\def \kdvs#1#2#3{Kong.~Danske Vid.~Selsk., Matt-fys.~Medd. #1 (#3) No.~#2}
\def \ky{Proceedings of the International Symposium on Lepton and
Photon Interactions at High Energy, Kyoto, Aug.~19-24, 1985, M.
Konuma and K. Takahashi, eds. (Kyoto Univ., Kyoto, 1985)} 
\def \latm{Lattice 1995 (Proceedings of the International Symposium on
Lattice Field Theory, Melbourne, Australia, 11--15 July 1995, T. D. Kieu,
B. H. J. McKellar, and A. Guttman, eds., North-Holland, Amsterdam, 1996}
\def \lgb{LP95: Proceedings of the International Symposium on Lepton and
Photon Interactions (IHEP), 10--15 August 1995, Beijing, People's Republic of
China, Z.-P. Zheng and H.-S. Chen, eds., World Scientific, Singapore, 1996} 
\def \lgg{International Symposium on Lepton and Photon Interactions, Geneva,
Switzerland, July, 1991}
\def \lkl87{Selected Topics in Electroweak Interactions (Proceedings of 
the Second Lake Louise Institute on New Frontiers in Particle Physics, 15--21
February, 1987), edited by J. M. Cameron \ite~(World Scientific, Singapore,
1987)}
\def \lti{lectures at this Institute}
\def \mpla #1#2#3{Mod.~Phys.~Lett. A #1 (#3) #2}
\def \nc#1#2#3{Nuovo Cim. #1 (#3) #2}
\def \nima#1#2#3{Nucl.~Inst.~Meth.~A #1 (#3) #2}
\def \np#1#2#3{Nucl.~Phys. #1 (#3) #2}
\def \npbps#1#2#3{Nucl.~Phys.~B (Proc.~Suppl.) #1 (#3) #2}
\def \npps#1#2#3{Nucl.~Phys.~(Proc.~Suppl.) #1 (#3) #2}
\def \oxf{Proceedings of the Oxford International Conference on
Elementary Particles 19/25 Sept.~1965, ed.~by T. R. Walsh (Chilton, Rutherford
High Energy Laboratory, 1966)}
\def \pascos{PASCOS 94 (Proceedings of the Fourth International
Symposium on Particles, Strings, and Cosmology, Syracuse University, 19--24
May 1994), ed.~by K. C. Wali (World Scientific, Singapore, 1995)}
\def \pbarp{AIP Conference Proceedings 357: 10th Topical Workshop on
Proton-Antiproton Collider Physics, Fermilab, May 1995, ed.~by R. Raja and J.
Yoh (AIP, New York, 1996)}
\def \pisma#1#2#3#4{Pis'ma Zh. Eksp. Teor. Fiz. #1 (#3) #2 [JETP Lett.,
#1 (#3) #4]} 
\def \pl#1#2#3{Phys.~Lett. #1 (#3) #2}
\def \pla#1#2#3{Phys.~Lett. A #1 (#3) #2}
\def \plb#1#2#3{Phys.~Lett. B #1 (#3) #2}
\def \ppmsj#1#2#3{Proc.~Phys.~Math.~Soc.~Jap. #1 (#3) #2}
\def \pnpp#1#2#3{Prog.~Nucl.~Part.~Phys. #1 (#3) #2}
\def \pr#1#2#3{Phys.~Rev. #1 (#3) #2}
\def \prd#1#2#3{Phys.~Rev. D #1 (#3) #2}
\def \prl#1#2#3{Phys.~Rev.~Lett. #1 (#3) #2}
\def \prp#1#2#3{Phys.~Rep. #1 (#3) #2}
\def \ptp#1#2#3{Prog.~Theor.~Phys. #1 (#3) #2}
\def \ptwaw{Plenary talk, in Proceedings of the XXVIII International Conference
on High Energy Physics, Warsaw, July 25--31, 1996, Z. Ajduk and A. K.
Wroblewski, eds. (World Scientific, River Edge, NJ, 1997)}
\def \rmp#1#2#3{Rev.~Mod.~Phys. #1 (#3) #2} 
\def \si90{25th International Conference on High Energy Physics, Singapore,
Aug. 2-8, 1990, Proceedings edited by K. K. Phua and Y. Yamaguchi (World
Scientific, Teaneck, N. J., 1991)}
\def \slaclg{Proceedings of the 1975 International Symposium on
Lepton and Photon Interactions at High Energies, Stanford University,
Aug.~21--27, 1975, W. T. Kirk, ed., SLAC, Stanford, CA, (1975)} 
\def \slc{Proceedings of the Salt Lake City Meeting (Division of
Particles and Fields, American Physical Society, Salt Lake City, Utah, 1987),
ed. by C. DeTar and J. S. Ball (World Scientific, Singapore, 1987)}
\def \smass{Proceedings of the 1982 DPF Summer Study on Elementary
Particle Physics and Future Facilities, Snowmass, Colorado, edited by R.
Donaldson, R. Gustafson, and F. Paige (World Scientific, Singapore, 1982)}
\def \smassa{Research Directions for the Decade (Proceedings of the
1990 DPF Snowmass Workshop), edited by E. L. Berger (World Scientific,
Singapore, 1991)}
\def \smassb{Proceedings of the Workshop on $B$ Physics at Hadron
Accelerators, Snowmass, Colorado, 21 June--2 July 1994, ed.~by P. McBride
and C. S. Mishra, Fermilab report FERMILAB-CONF-93/267 (Fermilab, Batavia, IL,
1993)} 
\def \stone{B Decays, edited by S. Stone (World Scientific, Singapore,
1994)}
\def \tp{these Proceedings}
\def \Vanc{XXIX International Conference on High Energy Physics, Vancouver,
BC, Canada, July 23--29, 1998, Proceedings}
\def \waw{XXVIII International Conference on High Energy
Physics, Warsaw, July 25--31, 1996, Proceedings edited by Z. Ajduk and A. K.
Wroblewski (World Scientific, River Edge, NJ, 1997)}
\def \yaf#1#2#3#4{Yad.~Fiz. #1 (#3) #2 [Sov.~J.~Nucl.~Phys.,~#1 (#3) #4]}
\def \zhetf#1#2#3#4#5#6{Zh.~Eksp.~Teor.~Fiz. #1 (#3) #2 [Sov.~Phys.~--
JETP, #4 (#6) #5]}
\def \zhetfl#1#2#3#4{Pis'ma Zh.~Eksp.~Teor.~Fiz. #1 (#3) #2 [JETP
Letters #1 (#3) #4]}
\def \zp#1#2#3{Zeit.~Phys. #1 (#3) #2}
\def \zpc#1#2#3{Zeit.~Phys.~C #1 (#3) #2}

\end{document}